# Terahertz in-line digital holography of dragonfly hindwing: amplitude and phase reconstruction at enhanced resolution by extrapolation

Lu Rong,[1] Tatiana Latychevskaia,[2] Dayong Wang,[1*] Xun Zhou,[3,4] Haochong Huang,[1] Zeyu Li,[3,4] and Yunxin Wang[1]

[1]*College of Applied Sciences, Beijing University of Technology, No. 100 Pingleyuan Rd., Beijing, 100124, China*
[2]*Physics Institute University of Zurich, Winterthurerstrasse 190, 8057 Zurich, Switzerland*
[3]*Reserach Center of Laser Fusion, CAEP, Mianyang, 621900, China*
[4]*Terahertz Research Center, CAEP, Mianyang, 621900, China*
[*]*wdyong@bjut.edu.cn*

## Abstract

We report here on terahertz (THz) digital holography on a biological specimen. A continuous-wave (CW) THz in-line holographic setup was built based on a 2.52 THz $CO_2$ pumped THz laser and a pyroelectric array detector. We introduced novel statistical method of obtaining true intensity values for the pyroelectric array detector's pixels. Absorption and phase-shifting images of a dragonfly's hind wing were reconstructed simultaneously from single in-line hologram. Furthermore, we applied phase retrieval routines to eliminate twin image and enhanced the resolution of the reconstructions by hologram extrapolation beyond the detector area. The finest observed features are 35 μm width cross veins.

## 1. Introduction

Terahertz electromagnetic waves have frequencies approximately between 0.1 and 10 THz with corresponding wavelengths between 0.1 and a few millimeters and can penetrate a wide variety of nonconducting materials that may be opaque to visible light and demonstrate low contrast when imaged with X-rays. On the other hand, terahertz waves are easily absorbed by conducting materials, like, for instance, water. These unusual properties make terahertz waves a good choice for imaging thick layered biological tissues. Moreover, terahertz radiation is nondestructive for biological specimens and can potentially be better than X-rays for imaging biological samples. For an overview of THz imaging for biomedical applications see [1].

However, due to the relatively long wavelength and the beam size compared to the wavelength, the diffraction effects play a serious role in focusing and imaging with terahertz waves. For this reason, optical elements such as lenses should be avoided. Imaging without any optical elements between the sample and the detector is possible by applying holography technique, when the scattered object wave is superimposed with a known reference wave, and the resulting interference pattern contains sufficient information about the complex-valued object wavefront to reconstruct the object [2-3].

For the first time, digital holography with terahertz radiation was realized in 2004 by Mahon et al. [4-5] using a 0.1 THz Gunn diode oscillator. The holograms were obtained in an off-axis scheme and recorded by a scanning detector. The smallest feature resolvable in the reconstruction was about 9 mm [5]. Terahertz holography with a free-electron laser was reported by Cherkassky et al. in 2005 [6] and later in 2010 by Knyazev et al. [7]. They recorded off-axis holograms using a combination of thermal image plate and intensified CCD camera. However, the short length of the FEL pulse provided limited coherence length and no reconstructions of experimental holograms were demonstrated. In 2011, Heimbeck et al. [8] demonstrated off-axis terahertz holography employing a highly coherent, frequency tuneable terahertz source composed of an 8–20 GHz Micro Lambda Wireless microwave synthesizer and a Virginia Diodes, Inc. multiplier chain. The holograms were recorded by a scanning Schottky diode detector and reconstructed using dual-wavelength reconstruction methods. In 2011, Ding et al. [9] employed an SIFIR-50 Coherent, Inc. terahertz laser, which could generate a 2.52 THz (wavelength is 118.83 μm) continuous-wave THz output. Their holograms were detected by means of a pyroelectric array detector and the reconstructions could resolve the smallest feature of about 0.4 mm. Li et al. in the same group [10] also made a comparison between the Fresnel algorithm and the angular spectrum algorithm in the THz domain. The same researchers reported in 2012 an improved reconstruction technique: this is achieved by upsampling of the off-axis holograms and suppressing the zero-order artifact, which allowed the resolution to be improved by up to 0.245 mm [11]. In 2012, the same group demonstrated for the first time THz Gabor in-line holography using the same source and detector [12-13]. The reconstructions demonstrated a resolution of about 0.2 mm. They also verified the feasibility of THz Gabor in-line digital holography in imaging concealed objects [12]. However, no simultaneous reconstruction of absorption and phase-shifting distributions from a single terahertz in-line hologram has yet been reported.

Off-axis and in-line holographic schemes have their advantages and disadvantages [14]. An off-axis scheme allows relatively easy reconstruction of a complex-valued wavefront, but provides limited resolution due to the required spectral filtering. An in-line scheme does not require an additional reference beam and provides a resolution that is only limited by the numerical aperture of the setup, but the reconstructions suffer from so-called twin image, the problem that was first described in original Gabor's works [2-3]. Modern reconstruction algorithms allow elimination of the twin image in reconstructions from a single in-line hologram [15-18] or two or more in-line holograms [19-22].

## 2. Sample

The sample was a dragonfly hindwing. The flight performance of dragonflies is remarkable due to the highly evolved and largely optimized mechanical construction of their wings. The unique structure of dragonfly wings has become a research focus, i.e., the thickness, wing shapes, venation pattern, surface roughness, and pterostigma position have been studied and exploited in various bionic, biomechanical and aerodynamic applications [23-25]. Zeng et al. were able to measure the venation shape and thickness of the wing membrane by employing a 633 nm laser light interferometer, but the thickness of the veins could not be measured because veins are opaque in visible light [23]. Jongerius et al. performed micro computed tomography, which allowed the wing profile to be obtained; however, the sample had to be dried for 12 hours before the acquisition, which deformed the wing [25]. Li et al. [24] and Ren et al. [26] reported scanning electron microscope (SEM) images of the vein profiles. From their SEM images, one concludes that different veins have different morphology. For example, the longitudinal veins are roughly 50 μm in diameter, tubular-shaped with "walls," having a sandwich-like microstructure, and containing chitin and protein. Moreover, cone-like spikes are observed on the surface of the longitudinal veins, the bottom diameter and height of which are 10 μm and 20 μm, respectively. Their purpose is to increase the surface roughness of veins, which surprisingly reduces the aerodynamic friction resistance similar to the effect of dimples in a golf ball [27]. Therefore, we will call the vein extension out of the wing plane the vein's "thickness," and the vein's width in the wing's plane the "width." THz digital holography can retrieve both the absorption and phase of the dragonfly hindwing. A digital optical microscopic image (50×) of the sample is shown in Fig. 1, where some veins are indicated. The width of the costa, radius vein, median veins and cross veins measured from the photograph are approximately 110 μm, 70 μm, 35–48 μm, and 33–40 μm, respectively.

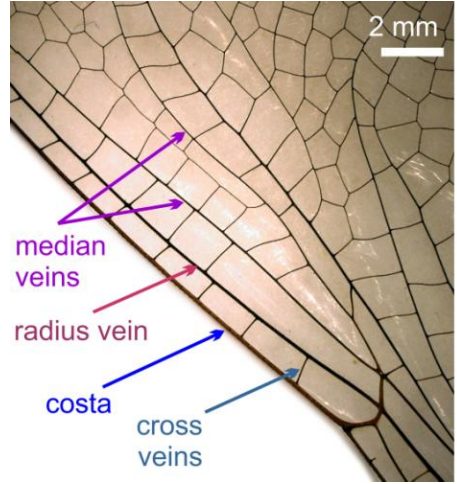

Fig. 1. Digital optical microscopic image (50×) of the sample dragonfly hindwing. The veins are assigned according to the map provided in [24].

### 3. Experimental setup

An experimental setup for terahertz in-line digital holography is depicted in Fig. 2. The source was an FIRL100, Edinburgh Instrument Ltd. terahertz laser: an FIR laser system pumped by a $CO_2$ laser. The operating frequency was 2.52 THz (wavelength is 118.83 μm) with output power of about 150 mW. The beam direction was changed by a 50 mm diameter gold-coated reflecting mirror (M). Next, the beam was expanded and collimated via two gold-coated off-axis parabolic mirrors (PM1 and PM2), the effective focal lengths of which were 76.2 mm and 152.4 mm, respectively. The hologram was formed by the interference between the wave scattered off the object and the nonscattered wave. A pyroelectric array detector (Pyrocam III, Ophir-Spiricon, Inc.) was used to record the holograms; it has 124 × 124 pixels with a pixel size of 85 μm × 85 μm and pitch of 100 μm × 100 μm, thus its total size is 12.4 mm × 12.4 mm. To detect the continuous-wave power, there was a frequency chopper enclosed inside the detector. In the experiment, the chopping frequency was set at 48 Hz.

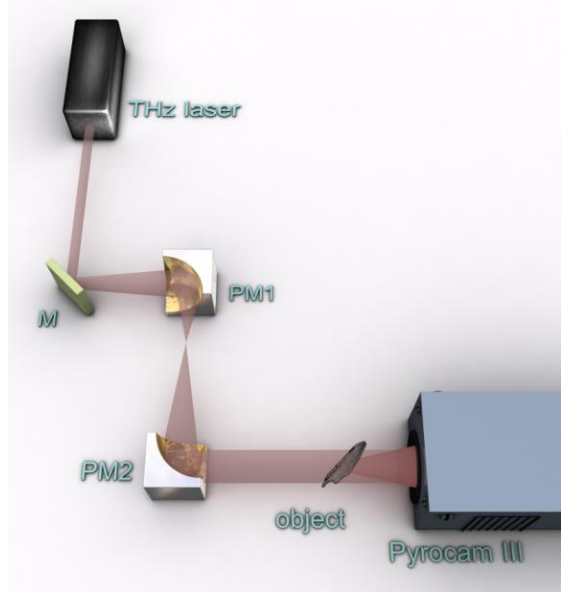

Fig. 2. Schematic layout of the terahertz in-line digital holography. For details see the main text.

## 4. Recording and pre-reconstruction steps

For simultaneous reconstruction of absorption and phase-shifting distributions of the object [16, 28], the hologram distribution $H_0(X,Y)$ must be divided by the background distribution $B_0(X,Y)$ recorded in the absence of the object; here $(X,Y)$ are the coordinates in the detector plane. It was very difficult to record terahertz in-line hologram and background images with a good signal-to-noise ratio for the following reasons: (1) The beam profile was difficult to optimize and the beam size was fixed by the focal length ratio of the parabolic mirrors. (2) The whole experimental arrangement had to be compact to reduce absorption of the THz beam by water vapor in the air. (3) The intensity of holograms periodically fluctuated due to the air vibration from the enclosed camera chopper. (4) The distance between the sample and the detector has to be small enough to guarantee a large enough diffraction angle. Thus, the sample had to be placed very close (almost touching) to the detector. To increase the signal-to-noise ratio in images, 1000 frames were recorded for both hologram $H_0(X,Y)$ and background (without object) $B_0(X,Y)$ images.

A pyroelectric detector, like the one employed in the experiments reported here, can deliver pixel values that are zero or negative [29]. The detected signal, in general, contains Gaussian distributed noise and thus negative values can occur at some pixels at some frames. To solve the issue with the negative pixels and obtain an average intensity at each pixel, we performed the following data treatment. The intensity at each pixel is extracted for each of the 1000 frames and the obtained one-dimensional intensity distribution is presented in the form of a histogram, where the width of the bars for N=1000 samples is chosen to be $N^{1/3}=10$ pixels, as illustrated in Fig. 3. The histogram is fitted with Gaussian distribution and its mean is selected to be the statistically averaged intensity at the pixel. The pixels whose values are zero for all 1000 frames are marked as "dead" pixels and excluded from the reconstruction by leaving their values free of constraints (for details see below).

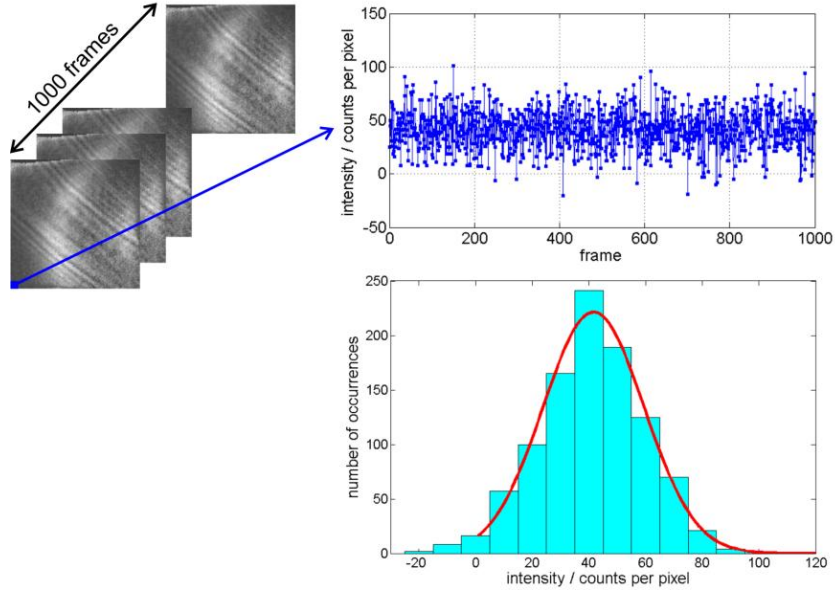

Fig. 3. Histogram of intensity values at pixel (1,1) in the sequence of 1000 hologram frames and its Gaussian fit with the center at 41.66 a.u.

The output power of the terahertz source was not constant during the experiment, and thus it appears that the averaged $H_0(X,Y)$ and $B_0(X,Y)$ were recorded at different intensities. As a consequence, the values of the normalized hologram $H_0(X,Y)/B_0(X,Y)$ do not vary around 1 (amplitude of the reference wave) at the regions where the object wave is much weaker than the reference wave. For instance, in the $60 \times 60$ pixels patch in the bottom left corner of the hologram (see Fig. 4), the averaged intensity amounts to $f = 1.12$ instead of 1. To equalize the hologram distribution to the reference wave amplitude level, the normalized hologram $H_0(X,Y)/B_0(X,Y)$ is divided by the factor $f$: $H(X,Y) = H_0(X,Y)/(fB_0(X,Y))$. The values of the dead pixels are replaced by the amplitude of the reference wave 1.

## 5. Angular spectrum reconstruction

In the experiments presented here, the distance between the sample and the detector is comparable to the sample size, therefore the Fresnel approximation $z^3 \gg \frac{\pi}{4\lambda}\left[(x-X)^2+(y-Y)^2\right]^2_{max}$ is not fulfilled; here $z$ is the distance between the sample and the detector, $\lambda$ is the wavelength, and $(x,y)$ and $(X,Y)$ are the coordinates in the object plane and the detector plane, respectively. For that reason, the angular spectrum theory [30], which describes the propagation of plane waves for short distances, is applied here for the reconstruction. The complex-valued transmission function of the sample $t(x,y)$ is reconstructed by the propagation of the optical field from the detector plane backwards to the object plane [30]:

$$t(x,y) = \mathrm{FT}^{-1}\left[\mathrm{FT}\{H(X,Y)\}\exp\left(-\frac{2\pi i z}{\lambda}\sqrt{1-(\lambda f_x)^2-(\lambda f_y)^2}\right)\right] \qquad (1)$$

where FT and FT$^{-1}$ are the Fourier transform and inverse Fourier transform, respectively, and $(f_x, f_y)$ are the spatial frequencies. The absorption $a(x,y)$ and phase $\phi(x,y)$ are then extracted using the following relation [28]:

$$t(x,y) = \exp\left[-a(x,y)-i\phi(x,y)\right] \qquad (2)$$

The absorption and phase distributions reconstructed using Eq.1–2 at a distance of 16.6 mm from the detector are shown in Fig. 4. The reconstruction shows the venation pattern of a dragonfly hindwing. The main veins, such as the costa and radius veins, are much more pronounced than the cross veins connecting them. In the top right corner of both reconstructed distributions, the reconstructed veins are superimposed with the twin image, and it is difficult to judge which stripes are veins and which are the fringes due to the twin image.

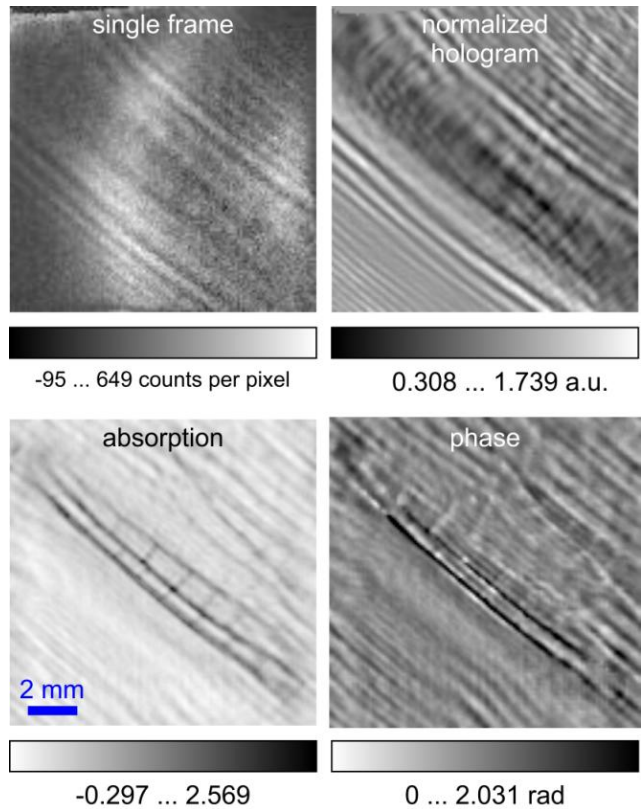

Fig. 4. Single frame and normalized hologram of dragonfly hindwing and the reconstructed absorption and phase distributions.

In order to see whether the fringes in the top right corner of the reconstructions are veins or the twin image, we also applied a single-sideband reconstruction method. Single-sideband holography [31] offers a simple tool for reconstructing an object without a twin image on one side of the object [32]. Single-sideband reconstruction is achieved by the following trick: During the reconstruction, after the hologram is Fourier transformed either the left or right side of the spectrum is set to zero, and as a result, the corresponding side of the object appears twin-image-free in the reconstruction. In our case, the hologram was first padded with zeros up to 1000×1000 pixels, then rotated by 45°, then single-sideband reconstructed, rotated back by -45°, and the center piece of 124×124 pixels was extracted. The resulting reconstructions are shown in Fig. 5. The bottom row in Fig. 5, where the reconstructions without twin image in the top right corner are shown, clearly demonstrates that the fine equidistant fringes are gone, and therefore those fringes were due to the twin image.

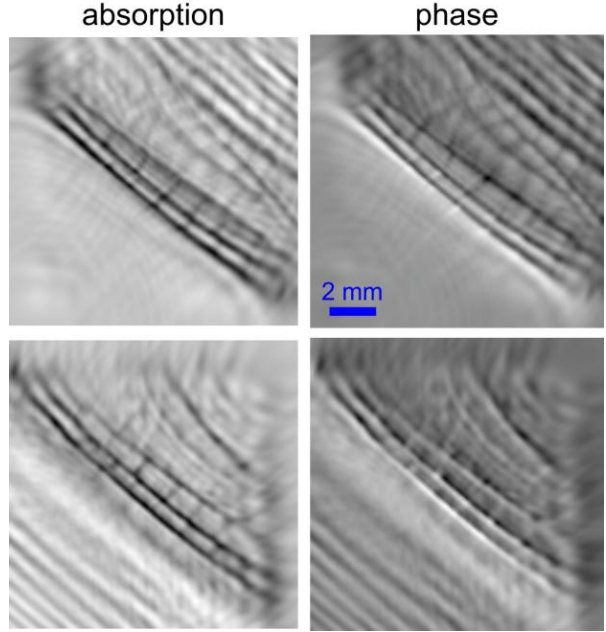

Fig. 5. Single-sideband reconstructions of a dragonfly hindwing. Left column: absorption distribution. Right column: phase distribution. Top row: Twin image is removed in the bottom left corner. Bottom row: Twin image is removed in the top right corner. (Because single-sideband reconstruction removes half of the spectrum, the quantitative information about absorption and phase distribution is lost, and the intensity bars are omitted to avoid misinterpretation.)

## 6. Iterative reconstruction

Iterative phase retrieval has previously been shown to successfully remove twin images in the reconstruction of in-line holograms from a single holographic record [14-18]. We applied an algorithm similar to the algorithms reported in [14, 16]. The positive absorption and mask support were used as the constraints in the object domain, while the magnitude of the normalized hologram was used as the constraint in the detector plane. The values of the "dead" pixels were set free of constraints and their values were updated during the iterative procedure.

Every 10$^{th}$ iteration, the reconstructed absorption distribution was smoothed by convolution with a 3 × 3 pixels filter in the form of a Gaussian distribution

$$K = \begin{pmatrix} 1 & 1 & 1 \\ 1 & 4 & 1 \\ 1 & 1 & 1 \end{pmatrix} \qquad (3)$$

to smooth and suppress the accumulation of noisy peaks. Before the iterative reconstruction the hologram was upsampled to 512×512 pixels to increase the resolution [33] and to be able to apply the smoothing filter over a larger area.

The resulting reconstructions after 1000 iterations are shown in Fig. 6. The twin image is suppressed and the venation pattern becomes apparent. The phase distribution has been additionally reconstructed at a plane slightly out of focus, as the phase distribution at a plane behind the object has a smoother, more unwrapped appearance than when it is reconstructed in the object plane.

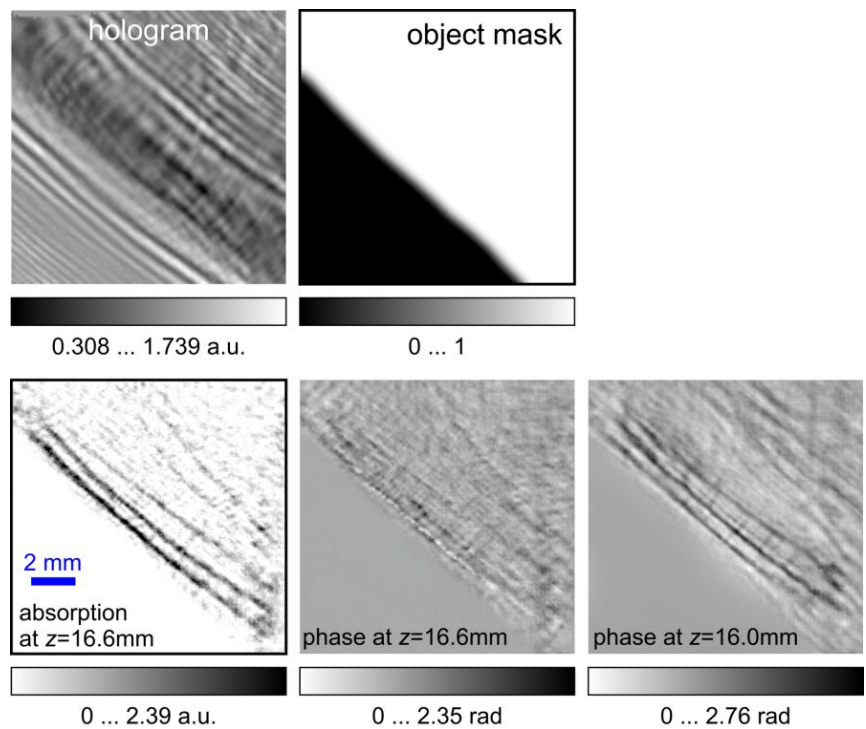

Fig. 6. Iterative reconstruction of a dragonfly hindwing. The hologram and masking support in the object domain are shown in the top row. The absorption and phase distribution reconstructed at z = 16.6 mm are shown in the bottom row. Additional reconstruction of phase distribution at z = 16.0 mm, i.e. slightly out of focus where the phase is unwrapped, is also shown.

## 7. Iterative reconstruction with extrapolation

The small size of the detector, or small numerical aperture (N.A.) of the setup, results in a limited resolution in the reconstructed images [34]. Recently, extrapolation of holograms has been proposed, which *a posteriori* increases the detector size, N.A., and hence, the resolution [35-36]. However, the extrapolation method requires the object to be isolated, so that the reference wave surrounds the object. The challenge of the present work is that the dragonfly hindwing is not an isolated object. We applied the same extrapolation procedure as described in detail in [35] but with a modified padding of the hologram. For the first iteration, the original hologram was padded up to 500 × 500 pixels with a background of varying amplitude to match the transmission of the hologram. As shown in Fig. 7, "padding," noise distributed in the range (-0.1,0.1), was added to the padding. During the iterative routine, in the object domain we applied the constraint of positive absorption, and no constraint was applied to the phase distribution. The results after 1000 iterations are shown in Fig. 7. The hologram has extrapolated itself beyond the original holographic record, but mostly in the bottom left corner, where the interference between the reference wave and the object wave is recorded. The extrapolation is weaker in the top right corner, where no clear reference wave is provided. Extrapolation fails at the edges of strong absorbing main veins (top left and bottom right corners). Even though the hologram has been extrapolated only partially, some improvements can already be seen when comparing the reconstructions of the extrapolated hologram with the reconstructions obtained by the twin image removal routine shown in Fig. 6. The cross veins are now better resolved and the median veins become more apparent. Both the absorption and phase reconstruction of the costa exhibit some peaks, which can be an indication of the spikes. However, the weak fringes due to the twin image are still present; this may be due to the very loose masking support in the object domain in the top right corner, shown in Fig. 7.

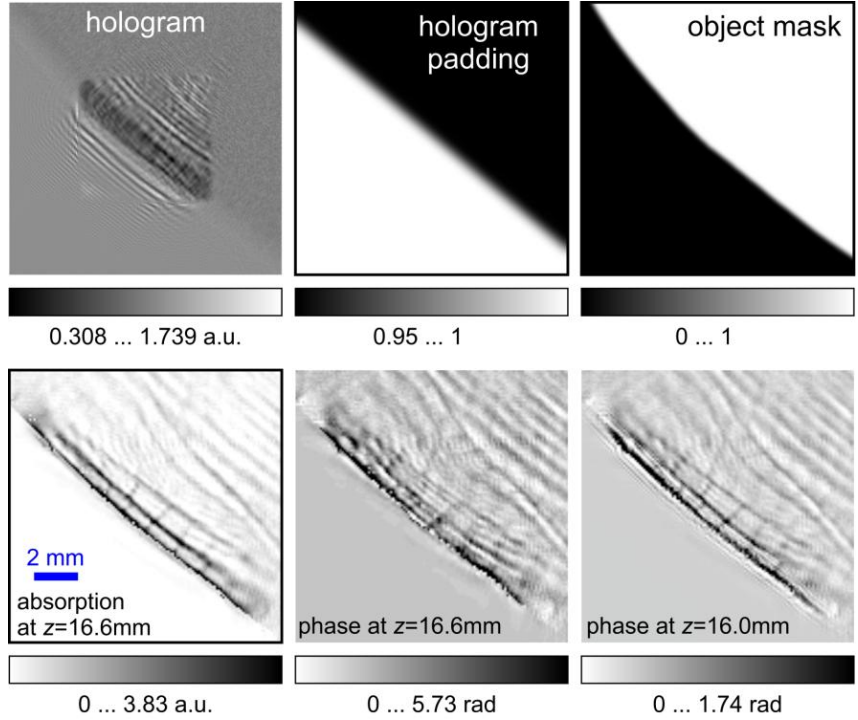

Fig. 7. Iterative reconstruction with extrapolation. 250 × 250 pixels fragment of the extrapolated hologram, hologram padding and object masking support are shown in the top row. The absorption and phase distribution reconstructed at z = 16.6 mm and additional reconstruction of phase distribution at z = 16.0 mm are shown in the bottom row.

## 8. Resolution

An in-line holographic setup allows for achieving the highest possible resolution, limited only by the size of the detector and the wavelength. The intrinsic resolution of the setup presented here, according to the Abbe criterion, amounts to $R = \frac{\lambda}{2\sin\vartheta_{max}} = 164.4$ μm. The extrapolation procedure increased the effective size of the screen from 12.4 mm × 12.4 mm to approximately 33.0 mm × 33.0 mm, and thus, the expected resolution is $R_{extrapolated} = 82.8$ μm. The smallest features resolved in the reconstruction are cross veins, compare Fig. 1 and Fig. 7. Their width in the reconstruction is about 1–2 pixels, or 124–248 μm, but the width estimated from the optical microscopic photograph is about 35 μm. Thus, features smaller than the wavelength are reconstructed. One possible explanation for this remarkable result is that the wavefront scattered even by the smallest features is distributed over the entire surface of the detector and therefore recorded in the hologram; the reconstruction therefore reveals such small features. However, should two such veins be closely spaced, they would not be resolved.

## 9. Conclusion

We performed continuous-wave terahertz in-line digital holography on a biological specimen: a dragonfly's hind wing. For the first time, terahertz holography was applied to an image of a biological specimen. Also for the first time, we demonstrated simultaneous reconstruction of absorption and phase-shifting distributions of the sample from single in-line terahertz hologram, that is, hologram recorded in an in-line holographic scheme that employs no optical elements between the sample and the detector. This allowed us to achieve the highest possible resolution. A novel method of the analysis of intensity values for pyroelectric array detector pixels is presented. This method is based on applying statistical analyses and assigning the statistical average as the true intensity value. An advanced method of iterative extrapolation is proposed, which allowed for the extrapolation of a hologram of the object that is not surrounded by known support. The contrast of the reconstructed absorption and phase-shifting distributions decreases from leading to trailing edge and is almost constant along the proximo-distal of longitudinal veins. The costa and radius veins demonstrate superior contrast when compared with other veins. The cross veins of a width of 35 μm can be observed in the reconstructions. In view of the unique character of a THz beam, CW-THz in-line digital holography has the potential to become a complementary imaging technique for biomedical applications. Further work would be dedicated to further improving the resolution of the reconstructed objects, which can be achieved by increasing the effective N.A. of the setup, which in turn implies either getting a larger detector or/and improving the algorithm for the extrapolation of holograms.

Authors' contributions: L. R., D. W., and X. Z. initiated the experiment, H. H., Z. L., and Y. W. built the experimental setup and conducted the data acquisition, T. L. developed the numerical routines and performed the reconstructions, T.L. and R. L. wrote the manuscript.

## Acknowledgments

This work is financially supported by the National Natural Science Foundation of China (No. 61307010 and 61205010), the Beijing Municipal Natural Science Foundation (No. 1122004), the China Postdoctoral Research Foundation (No. 2013M540828), the Beijing Postdoctoral Science Foundation (No. 2013ZZ-17), the Research Fund for the Doctoral Program of Higher Education of China (No. 20121103120003), and the Beijing University Fund for Scientific Research of Doctor (No. X0006111201102). The Swiss National Science Foundation (No. 140764) and the University of Zurich are gratefully acknowledged for their financial support.